\documentstyle[12pt]{article}
\newcommand {\dis}{\displaystyle}
\newcommand {\beq}{\begin{eqnarray}}
\newcommand {\eeq}{\end{eqnarray}}
\newcommand {\beqs}{\begin{eqnarray*}}
\newcommand {\eeqs}{\end{eqnarray*}}
\newcommand {\del}{\partial}
\newcommand {\delz}{z\frac{d}{dz}}
\newcommand {\ca}{\cal A}
\newcommand {\ct}{\cal T}

\newcommand {\bc}{\bf C}
\newcommand {\bz}{\bf Z}
\newcommand {\ia}{\it A}
\newcommand {\itt}{\it T}
\newcommand \zg[1]{{Z}^{\gamma}_{#1}}

\newcommand {\vpi}{\varpi}
\newcommand \vvp[1]{{\varpi}_{#1}}

\newcommand {\Om}{\Omega}
\newcommand \om[1]{{\omega}_{#1}}
\newcommand \gam[1]{\Gamma \left({#1}\right)}
\newcommand \gamm[2]{\Gamma \left({\frac{#1}{#2}}\right)}

\newcommand {\mfd}{\mbox{manifold}}
\newcommand {\mfds}{\mbox{manifolds}}
\newcommand {\md}{\mbox{moduli space}}
\newcommand {\mds}{\mbox{moduli spaces}}
\newcommand {\cs}{\mbox{complex structure}}
\newcommand {\css}{\mbox{complex structures}}
\newcommand {\ks}{\mbox{K{\"a}hler structure}}
\newcommand {\kss}{\mbox{K{\"a}hler structures}}
\newcommand {\kae}{\mbox{K{\"a}hler}}
\newcommand \com[2]{\raisebox{-0.325ex}{\scriptsize {#1}} C_{#2}}
\newcommand \hyp[2]{\raisebox{-0.325ex}{\scriptsize {#1}} F_{#2}}
\newcommand \thyp[2]{\raisebox{-0.325ex}{\scriptsize {#1}}{\tilde{F}}_{#2}}

\newcommand \fsum[1]{\sum_{#1}^{\infty}}

\newcommand {\mm}{\cal M}
\newcommand {\mw}{\cal W}
\newcommand{\pr}{\hspace{\parindent}}

\begin{document}
\setlength{\oddsidemargin}{0cm}
\setlength{\baselineskip}{7mm}

\begin{titlepage}
    \begin{normalsize}
     \begin{flushright}
             {\bf UT-663} \\
             November 1993
     \end{flushright}
    \end{normalsize}
    \begin{LARGE}
       \vspace{1cm}
       \begin{center}
        {Mirror Symmetry\\
              of \\
          K3 and Torus } \\
       \end{center}
    \end{LARGE}

   \vspace{5mm}

\begin{center}
           Masaru N{\sc agura}\footnote{A Fellow of the Japan Society
           for the Promotion of Science for Japanese Junior
           Scientists}{${}^{,}$}\footnote{E-mail address:
           nagura@tkyvax.phys.s.u-tokyo.ac.jp}\\
       \vspace{4mm}
             { \it and }\\
       \vspace{4mm}
           Katsuyuki S{\sc ugiyama}\footnote{E-mail address:
             ksugi@tkyvax.phys.s.u-tokyo.ac.jp} \\
       \vspace{4mm}
                  {\it Department of Physics, University of Tokyo} \\
                  {\it Bunkyo-ku, Tokyo 113, Japan} \\
       \vspace{1cm}

     \begin{large} ABSTRACT \end{large}
\par
  \end{center}
\begin{quote}
\begin{normalsize}

We discuss a K3 and torus from view point of "mirror symmetry".
We calculate the periods of the K3 surface
and obtain the mirror map, the two-point correlation
function, and the prepotential.
Then we find there is no instanton correction on K3 (also torus), which is
expected from view point of Algebraic geometry.

 \end{normalsize}
\end{quote}

\end{titlepage}
\vfill

\section{Introduction}

\pr
It is a long time since the string theory attract the attention of physicist
 as a candidate of the most fundamental theory
that explain all the physical phenomena in nature .
In spite of the intense research activities,
it remains an open problem to formulate these theories
even now. One of the difficulties seems to stem from the treatment of
the Calabi-yau spaces that should be compactified. In view of the particle
physics, (co)homology classes of this Calabi-yau spaces correspond to
zero mass fields in the low energy effective theory and this spaces
play very important roles in researching the properties of
the field contents, Yukawa couplings and so on
{\cite{CHSW,SW,CO}}.
In view of mathematics they correspond to the tangent spaces of the moduli
space
; the complex structure or the {\kae} structure.
It seemed hard to investigate these properties because of effects of
quantum corrections {\cite{DSWW,DG}}.

Recently great progresses have been made in the
understanding of the moduli space in Calabi-Yau manifolds by the
discovery of the new notion, called the "mirror symmetry"
{\cite{MIR4}}. It became
possible to extract the properties of the spaces.
Especially some Yukawa couplings correspond to the complex moduli
recieve no quntum correction either in loop or by instanton,
owing to the nonrenormalization theorem.
This suggests the new concept "quantized algebraic varieties" and
attracts physical and
mathematical attentions because of the close relations between
string theories and algebraic geometries. For that reason, it is
not too much to say that this shed new light on the algebraic geometry.
When one discusses the mirror symmetries, the (complex)
dimension of the Calabi-Yau {\mfds} is
restricted to be three so far . Nevertheless generalized mirror
{\mfds} with other dimensions are expected to be much
richer in contents and to be fascinating objects {\cite{DIV}}.

The aim of this paper is to construct mirror {\mfds} paired with the
(complex one dimensional) torus, the K3 surface and to investigate their
properties as a first attempt to the generalization in other dimensions.
Firstly we construct mirror {\mfds} by the method of
the orbifoldization .
That is, one operates a discrete symmetry which
the original {\mfd} has on it and removes singular points on this
and obtains a mirror {\mfd}.
Primarily we study deformations of the Hodge structures
which the original {\mfd} or its mirror {\mfd} have .
The information on the Hodge structures is encoded in periods of these
{\mfds}.
Therefore one can extract properties of their {\mds} associated
with the deformations of the {\css} or {\kss} by examining these periods
concretely.

In the past, it was difficult to study the Yukawa couplings
correspond to the {\kae} {\mds} directly
because of the quantum corrections on this spaces
in contrast with the complex {\mds}
which receive no quantum correction. In the case of the Calabi-Yau {\mfds},
however, one can correlate the {\kae}
{\md} of the original {\mfd} with the complex {\md} of its mirror {\mfd}.
To be more precise, we map the complex {\md} of the mirror {\mfd}
to the {\kae} {\md} of the original one and get the information on
the {\kae} {\mds} of the original {\mfd} exactly.
Especially we are interested in couplings of (co)homology elements.
Using the mirror map, we identify the Yukawa coupling  of the complex moduli
on the mirror manifold with that of the {\kae} moduli on the original manifold.
And we find there is no instanton correction on the K3 (and torus).

This paper is organized as follows. In section $2$, we construct
a mirror {\mfd} paired with a K3 surface,
by orbifoldization .
In section $3$, we explain Picard-Fuchs equations that
periods of the algebraic varieties satisfy. We write down
solutions of the Picard-Fuchs equation for the K3 case and
construct a mirror map. In section $4$, a coupling of
(co)homology elements is calculated. The result is suggestive of the
nonexistence of the rational curve on the K3 surface. In section $5$,
monodromy matrices associated with solutions of Picard-Fuchs equation
are obtained. Further a fundamental region of the monodromy group is
written down. Section $6$ is devoted to conclusions and comments.
\\\\
\section{Construction of mirror \mfd}

\pr
In this section, we take a K3 surface as an original {\mfd} $\mm$
and construct its mirror {\mfd} $\mw$ by orbifoldizing
the original {\mfd} $\mm$, and resolving its singularities.

\subsection[Construction of mirror \mfd ]{Orbifoldization}

\pr
We take a K3 surface which is represented by a defining equation
with order $4$ in the projective
space ${\bf CP}^3$:
\[
{\mm} \hspace{1.5mm}; \hspace{1.5mm}
 p={X_1}^{4} +{X_2}^{4} +{X_3}^{4} +{X_4}^{4} =0
\]
Hodge numbers of this variety $\mm$, $h_{p,q} ={dim}_{\bc}
{\cal H}_{p,q} (\mm , {\bz})$ are known and illustrated
in the following Hodge
diamond {\cite{GH}}:
\[
\begin{array}{cccccc}
  &   &  1 &   &   & \\
  & 0 &    & 0 &   & \\
1 &   & 20 &   & 1 & \\
  & 0 &    & 0 &   &\\
  &   &  1 &   &   &.

\end{array}
\]

That is to say, non-zero numbers read,
\beqs
&& {h_{0,0}}={h_{0,2}}={h_{2,0}}={h_{2,2}}=1\,\,,\\
&& {h_{1,1}}=20\,\,.
\eeqs
Nineteen cycles out of independent twenty cycles in the Dolbeault
(co)homology ${\cal H}_{1,1} (\mm)$ can be expressed in monomials of $X_i$.
The remaining one cycle is associated with the {\ks}
of {$\mm$} {\cite{SEI,LG}}.
Also the {\mfd} $\mm$ has discrete symmetry $G={{\bz}_{4}^{\otimes 2}}$,
\beqs
&&(1,0,0,3) \,\,\,, \\
&&(0,1,0,3) \,\,\,, \\
&&(0,0,1,3) \,\,\,,
\eeqs
where the symbol $({{n}_{1}},{{n}_{2}},{{n}_{3}},{{n}_{4}})$ means
that the {\mfd} $\mm$ is invariant under the transformation,
\beqs
&& ({{X}_{1}},{{X}_{2}},{{X}_{3}},{{X}_{4}}) \rightarrow
({\alpha}^{n_1} {X_1},{\alpha}^{n_2} {X_2},
{\alpha}^{n_3} {X_3},{\alpha}^{n_4} {X_4}) \,\,\,, \\
&& ({\alpha}={e^{\frac{\pi}{2} i}}=i) \,\,\,.
\eeqs
Furthermore these three transformations are not independent because
successive actions of these transformations result in
$(1,1,1,1)=(0,0,0,0)$ in the ${\bf CP}^{3}$.

Following the recipe in the case of the quintic Calabi-Yau {\mfd}
{\cite{MIR4}},
we construct a mirror {\mfd} $\mw$.
Firstly we operate the original {\mfd} with this symmetry $G$.
There are $24 (=4 \times \com{4}{2})$ singular points under this operation.
So we have to remove these points from the {\mfd} $\mm$ and blow up each point.
A {\mfd} $\mw$ obtained in this recipe is a mirror {\mfd} of $\mm$. Its
Euler number is calculated, as follows;
\[
\chi ({\mw})=\frac{24-24}{4^2} + \frac{4 \times 24}{4} =24 \,\,\,.
\]
In short,
\[
\chi ({\mw})=\chi ({\mm})=24 \,\,\,,
\]
and the mirror {\mfd} $\mw$ has the same Hodge diamond as $\mm$.
This reflects a self-duality of the K3 surface.
Note that there is one invariant monomial, ${X_1}{X_2}{X_3}{X_4}$
in the (co)homology class ${\cal H}_{1,1}$ under the action of $G$.
Since the $\mw$ is invariant under $G$, one can identify the monomial
${X_1}{X_2}{X_3}{X_4}$ as an element of ${\cal H}_{1,1}({\cal W})$.
We assume this monomial on the mirror manifold ${\cal W}$ corresponds
to the {\kae} class of the original ${\cal M}$ and
we can deform the {\cs}
in the mirror $\mw$ with this monomial. That is to say,
we consider one parameter family of a mirror {\mfd} ${\mw}_{\psi}$,
\beq
{\mw}_{\psi} \hspace{1mm} \mbox{;} && \{{p=0}\}/G -4\psi {X_1}{X_2}{X_3}{X_4}
 \,\,\,,\hspace{3mm} (\psi \in \bc) \nonumber \\
&=& \{{{p_{\psi}}=0}\} /G \,\,\,, \nonumber \\
{{p}_{\psi}} &:=& p-4\psi {X_1}{X_2}{X_3}{X_4} \nonumber \\
&=& {{X_1}^{4}}+{{X_2}^{4}}+{{X_3}^{4}}
 +{{X_4}^{4}}-4\psi {X_1}{X_2}{X_3}{X_4} \label{eqn:de}\,\,\,.
\eeq
The parameter $\psi$ is a coordinate of the {\cs} of the mirror {\mfd} $\mw$.
Using some map, called a mirror map, we can translate
 the information on the complex moduli in $\mw$
 into that on the {\kae} moduli on $\mm$.
In later section $4$, we explain these properties in more detail.

\section{Picard-Fuchs equation and mirror map}

\pr
In this section, we introduce periods of the mirror {\mfd} and write
down a differential equation, called a Picard-Fuchs equation
satisfied by the periods {\cite{GP,LER}}.
We can obtain solutions of this equation instead of carrying out the integral
of the periods and define a mirror map of
the K3 surface, which connects the {\cs} of $\mw$ with the {\ks}
of $\mm$.

\subsection[Picard-Fuchs equation and mirror map]{Holomorphic form and periods}

\pr
The structure of the {\md} of a Calabi-Yau {\mfd} is described
by giving the periods of its {\mfd} {\cite{GP}},
which is expressed by
a holomorphic form. Let us consider a $d$-dimensional
hypersurface $V$ defined by the following homogeneous polynomial
$W$ in ${\bf CP}^{d+1}$,
$$W:={X_1}^{d+2}+{X_2}^{d+2}+ \cdots +{X_{d+1}}^{d+2}+{X_{d+2}}^{d+2}$$
It is known that $V$ defined by $W$ is a Calabi-Yau manifold.
Then there is a holomorphic d-form $\Om$ on $V$ defined globally
and nowhere zero.
$\Om$ can be expressed by, {\cite{GP,ATI,HOL}}
\beqs
\Om &:=& \int_{\gamma} \frac{1}{W} \sigma \,\,\,, \\
\sigma &:=& \sum_{a=1}^{d+2} {(-1)}^{a}
 {X_a}{{dX}_1} \wedge \cdots \wedge \widehat{{dX}_a} \wedge \cdots
 \wedge {{dX}_{d+2}} \,\,\,,
\eeqs
where $\gamma$ is a small one-dimensional curve winding around the $V$.
The $(d+1)$-form $\sigma$ is invariant under discrete transformations
${\bz}_{d+2}^{\otimes d}$,
\beqs
& & ({X_1},{X_2},\cdots ,{X_{i-1}},{X_{i}},{X_{i+1}},\cdots ,
{X_{d+1}},{X_{d+2}}) \\
&\rightarrow & ({X_1},{X_2},\cdots ,{{X}_{i-1}},
{{\alpha}_{d+2}}{{X}_{i}},{X_{i+1}},\cdots ,
{X_{d+1}},{{\alpha}_{d+2}^{d+1}}{X_{d+2}}) \,\,\,, \\
& &{ \alpha}_{d+2} \equiv \exp {\frac{2\pi i}{d+2}}\,\,\,.
\eeqs
The set of periods are defined to be
the integral of $\Om$ over homology cycles ${\tilde{\Gamma}}_{\alpha}$
of $V$,
\beqs
{\widehat{\vpi}}_{\alpha} &=& \int_{{\tilde{\Gamma}}_{\alpha}} \Om \,\,\,,\\
{\tilde{\Gamma}}_{\alpha} &\in & {\cal H}_{d}(V) \,\,\,.
\eeqs
It is known that if we consider the following series,
\[
W^{(N)} = \sum_{i=1}^{N} {{x}_{i}}^{N}
-N\psi \prod_{i=1}^{N} {{x}_{i}} \hspace{7mm} (N\geq 3) \,\,\,,
\]
the differential equations satisfied by the periods corespond to
$1$, ${\dis \prod_{i=1}^N} x_i$, $({\dis \prod_{i=1}^N} x_i)^2$, $\cdots$,
$({\dis \prod_{i=1}^N} x_i)^{N-2}$
can be written {\cite{LER}},
\beqs
\biggl[{{\left({\delz}\right)}^{N-1} \hspace{-3mm}-z
\left({\delz +\frac{1}{N}}\right)\cdots
\left({\delz +\frac{N-1}{N}}\right)}\biggr] {\vpi}^{(N)} =0 \,\,\,,
\eeqs
where ${\vpi}^{(N)} :=-N\psi {\widehat{\vpi}}^{(N)}$
and $z:= {\psi}^{-N}$.
These equations are called Picard-Fuchs equations.

\subsection[Picard-Fuchs equation and mirror map]
{Application to Kummer surface}

\pr
When we apply the formulae for $N=4$, we can obtain the Picard-Fuchs
equation for the mirror {\mfd} of the K3 surface $\mw$
represented by the defining equation (\ref{eqn:de}),
\beq
&& \biggl[{{\left({\delz}\right)}^{3} \hspace{-2.5mm}-z
 \left({\delz +\frac{1}{4}}\right)
 \left({\delz +\frac{2}{4}}\right)
 \left({\delz +\frac{3}{4}}\right)}\biggr] {\vpi} =0 \label{eqn:pf3}\,\,\,,\\
&& z:={\psi}^{-4} \,\,\,,\nonumber
\eeq
We should note that this equation is the same as that of the original manifold
${\cal M}$. The periods of mirror are the same as the corresponding periods of
the original up to multiplication factor ${1/4}^2$.
We think this fact is a direct result from mirror symmetry.

Firstly let us study the property of solutions of this
differential equation around a point $z=0$.
This point $z=0$ is a regular singular point {\cite{MOR}}. In addition, all the
coefficients of the terms $\left({\delz}\right)^{n} \,\,,\hspace{1mm}(n=0,1,2)$
vanish at this point. So solutions of this equation have unipotent
monodromy of index $3$, i.e. maximally unipotent monodromy {\cite{MOR}}.
Independent solutions of equation (\ref{eqn:pf3}) can be solved,
\beq
{W_{1}}({\psi}) &=& \fsum{n=0} \frac{(4n)!}
{{(n!)}^{4} {(4\psi)}^{4n}} \label{eqn:w1} \,\,\,,\\
{W_{2}}({\psi}) &=& -4\times {W}_{1} \cdot \log (4\psi) \nonumber \\
 &+& 4\times \fsum{n=0} \frac{(4n)!}{{(n!)}^{4} {(4\psi)}^{4n}}
 \left[{\, \Psi (4n+1)-\Psi (n+1) \,}\right] \label{eqn:w2} \,\,\,,\\
{W_{3}}({\psi}) &=& {4^2}\times  W_{1} \cdot
{\left\{{\log (4\psi)}\right\}}^2  \nonumber \\
&-& 2 \cdot {4^2}\times \fsum{n=0} \frac{(4n)!}{{(n!)}^{4} {(4\psi)}^{4n}}
 \left[{\,\Psi (4n+1)-\Psi (n+1)\,}\right] \cdot \log (4\psi) \\
&+& {4^2}\times \fsum{n=0} \frac{(4n)!}{{(n!)}^{4} {(4\psi)}^{4n}}
 \biggl\{{{\left[{\,\Psi (4n+1)-\Psi (n+1)\,}\right]}^{2}
 +{\Psi}' (4n+1)-\frac{1}{4}
{\Psi}' (n+1)}\biggr\} \label{eqn:w3}\,\,\,. \nonumber
\eeq
As another expression, we can take a basis of this equation {\cite{LER}},
\beq
{\vpi}_{a} :=\frac{1}{z \cdot {(z-1)}^{1/2}}
\left({\frac{dz}{ds}}\right)
\cdot {s}^{a} \hspace{5mm} (a=0,1,2) \label{eqn:vpi} \,\,\,.
\eeq
In this formula, $s$ is a function of $z$ and is a solution of
the schwarzian differential equation,
\beq
\left\{{s,z}\right\} =\frac{1}{2} \cdot \frac{1}{z^2}
 +\frac{3}{8} \cdot \frac{1}{{(1-z)}^2}
 +\frac{13}{32} \cdot \frac{1}{z(1-z)} \label{eqn:sch} \,\,\,.
\eeq
It can be represented by a triangle function {\cite{ERD}},
\[
s\biggl({0,\frac{1}{4} ,\frac{1}{2} ;z}\biggr) \,\,\,.
\]
The $s(z)$ is also rewritten as a ratio of two independent solutions
${{y}_{\mbox{${\scriptstyle I}$}}}$, ${{y}_{\mbox{${\scriptstyle I\! I}$}}}$
of Gauss's hypergeometric differential equation {\cite{ERD}},
\beq
&& z(1-z) \frac{{d^2} y}{{dz}^2} +\left({1-\frac{3}{2} z}\right)
\frac{dy}{dz} -\frac{3}{64} y=0 \label{eqn:gauss} \,\,\,,\\
&& s=\frac{{{y}_{\mbox{${\scriptstyle I\! I}$}}}}
{{{y}_{\mbox{${\scriptstyle I}$}}}} \label{eqn:s} \,\,\,.
\eeq
Obviously solutions of (\ref{eqn:vpi}) satisfy a relation,
\beq
{{\vpi}_{0}}{{\vpi}_{2}}-{{\vpi}_{1}}^2 =0 \label{eqn:so1} \,\,\,.
\eeq
To be sure, there is arbitrariness in the choice of two
independent solutions ${{y}_{\mbox{${\scriptstyle I}$}}}$,
${{y}_{\mbox{${\scriptstyle I\! I}$}}}$ resulting from the relation,
\beq
&& \left\{{{\frac{As+B}{Cs+D}},z}\right\}
 =\bigl\{{s,z}\bigr\} \label{eqn:arb} \,\,\,,\\
&& \mbox{with} \hspace{3mm} AD-BC
 \neq 0, \hspace{3mm} (A,B,C,D \in {\bc}) \,\,\,.\nonumber
\eeq
Let us investigate this arbitrariness in the choice of $s$.
If one replaces $s$ with $\frac{\displaystyle As+B}{\displaystyle Cs+D}$,\,
$(AD-BC\neq 0\,;\, A,B,C,D \in {\bc})$, the solutions ${\vpi}_{a}$ transform
linearly,
\begin{equation}
\left(
\begin{array}{c}
{\vpi}_0 \\
{\vpi}_1 \\
{\vpi}_2
\end{array}
\right)
\rightarrow \frac{1}{AD-BC}
\left(
\begin{array}{ccc}
D^2 & 2CD & C^2 \\
BD  & AD+BC & AC \\
B^2 & 2AB & A^2
\end{array}
\right)
\left(
\begin{array}{c}
{\vpi}_0 \\
{\vpi}_1 \\
{\vpi}_2
\end{array}
\right) \label{eqn:tra1}\,\,\,.
\end{equation}
Even after this transformation, the relation (\ref{eqn:so1}) remains invariant.
In order to research this transformation matrix in detail,
let us introduce a new basis of the solution of (\ref{eqn:gauss}),
\begin{equation}
\left(
\begin{array}{c}
\om{0} \\
\om{1} \\
\om{2}
\end{array}
\right)
\equiv \frac{1}{2}
\left(
\begin{array}{ccc}
1 & 0 & 1 \\
0 & 2 & 0 \\
1 & 0 & -1
\end{array}
\right)
\left(
\begin{array}{c}
{\vpi}_0 \\
{\vpi}_1 \\
{\vpi}_2
\end{array}
\right) \,\,\,.
\end{equation}
Then the relation (\ref{eqn:so1}) is re-expressed,
\[
0=-{{\vpi}_0}{{\vpi}_2}+{{\vpi}_1}^{2} =
-{{\om{0}}^2}+{{\om{1}}^2}+{{\om{2}}^2} \,\,\,. \label{eqn:so2}
\]
On this new basis, the transformation (\ref{eqn:tra1}) can be written,
\begin{equation}
\left(
\begin{array}{c}
\om{0} \\
\om{1} \\
\om{2}
\end{array}
\right)
\rightarrow {\bf M}
\left(
\begin{array}{c}
\om{0} \\
\om{1} \\
\om{2}
\end{array}
\right) \,\,\,,
\end{equation}

\begin{equation}
{\bf M}=\frac{1}{AD-BC}
\left(
\begin{array}{crc}
\frac{1}{2}({A^2}+{B^2}+{C^2}+{D^2}) & AB+CD &
\frac{1}{2}(-{A^2}+{B^2}-{C^2}+{D^2}) \\
AC+BD & AD+BC & -AC+BD \\
\frac{1}{2}(-{A^2}-{B^2}+{C^2}+{D^2}) & -AB+CD &
\frac{1}{2}({A^2}-{B^2}-{C^2}+{D^2})
\end{array}
\right) \label{eqn:tra2} \,\,\,.
\end{equation}
Because a bilinear form $(-{{\om{0}}^2}+{{\om{1}}^2}+{{\om{2}}^2})$ is
invariant under the transformation (\ref{eqn:tra2}), the transformation
matrix ${\bf M}$ is an element of the group $SO(2,1)$.

Taking notice of the relation (\ref{eqn:so1}), we can decompose
the holomorphic $(2,0)$-form $\Om$ with a basis of homology cycles
$ \alpha , \beta , \gamma$, which belong to ${\cal H}_{2} ({\mw})$,
and the solutions (\ref{eqn:vpi}) into
\beq
{\Om}={{\vpi}_0} \alpha +{{\vpi}_1} \beta -{{\vpi}_2}
\gamma \label{eqn:ome} \,\,\,,
\eeq
such that these cycles satisfy relations,
\beq
&& {\int}_{\mw} {\alpha} \wedge {\gamma}=
{\int}_{\mw} {\gamma} \wedge {\alpha}=1 \,\,\,,\\ \nonumber
&& {\int}_{\mw} {\beta} \wedge {\beta}=2 ,\,\,\,\,
{\int}_{\mw} others=0 \label{eqn:cyc} \,\,\,. \nonumber
\eeq
Then the coupling of the holomorphic $2$-form $\Om$ with itself vanishes
obviously,
\[
{\int}_{\mw} {\Om} \wedge {\Om} =0\,\,\,.
\]
We can easily return to the relation (\ref{eqn:so1}) from (\ref{eqn:ome})
(\ref{eqn:cyc}).
The existance of these homology cycles; ${\alpha},{\beta}$ and ${\gamma}$
satisfying the above relations on $K3$ surface has been known in Algebraic
geometry.
Of course, the choice of the homology basis; ${\alpha},{\beta}$ and ${\gamma}$
is not unique. If ${\alpha}',{\beta}'$ and ${\gamma}'$ consist another basis
then they are related to
${\alpha},{\beta}$ and ${\gamma}$
by a matrix $N$;
\beq
\left(
     \begin{array}{ccc}
     {\alpha}'\\
     {\beta}'\\
     {\gamma}'
     \end{array}
\right)=N\left(
     \begin{array}{ccc}
     {\alpha}\\
     {\beta}\\
     {\gamma}
     \end{array}
\right)\,\,\,,
\eeq
where the matrix $N$ takes value in integer and preserve a matrix
$
\left(
\begin{array}{ccc}
0& 0& 1\\
0& 2& 0\\
1& 0& 0
\end{array}
\right)
\;\;
$ is invariant. In short, the following relation is satisfied.
\beq
N\left(
\begin{array}{ccc}
0& 0& 1\\
0& 2& 0\\
1& 0& 0
\end{array}
\right) N^{-1}=
\left(
\begin{array}{ccc}
0& 0& 1\\
0& 2& 0\\
1& 0& 0
\end{array}
\right) \;\;.
\eeq

\subsection[Picard-Fuchs equation and mirror map]{Mirror map}
\pr
Following the recipe of D.R.Morrison {\cite{MOR}}, we define the "mirror map
$t({\psi})$" by demanding a property $t({{e^{2\pi i}}z})=t(z)+1$,
\[
t(z):=\frac{1}{2\pi i} \cdot \frac{{W_2}(z)}{{W_1}(z)}\,\,\,.
\]
In the above formula, ${W_1}(z), {W_2}(z)$ are solutions of
the equation (\ref{eqn:pf3}) and are
given in (\ref{eqn:w1}) and (\ref{eqn:w2}).
Remarkably, these solutions can also be written as the
products of two solutions of a Gauss's
hypergeometric equation (\ref{eqn:gauss}).
\beq
W_1 &=& {y_1} \cdot {y_1} \label{eqn:gw1} \,\,\,,\\
W_2 &=& {y_1} \cdot {y_2}+\sqrt{2} \, \pi {y_1}
\cdot {y_1} \label{eqn:gw2} \,\,\,.
\eeq
The functions ${y_1}, {y_2}$ are defined,
\beq
&& {y_1}:= \hyp{2}{1} \!\left({\frac{1}{8},
 \frac{3}{8},1;z }\right) \label{eqn:y1} \,\,\,,\\
&& {y_2}:= \hyp{2}{1} \!\left({\frac{1}{8},
 \frac{3}{8},1;z }\right) {\log z}+{{\thyp{2}{1}}} \left({\frac{1}{8},
 \frac{3}{8},1;z }\right) \label{eqn:y2} \,\,\,,
\eeq
\beqs
{\tilde{\hyp{2}{1}}}
 \left({\frac{1}{8},\frac{3}{8},1;z }\right)
&=& \fsum{n=0} {\displaystyle
 \frac{ \gam{n+ \frac{1}{8}} \gam{n+ \frac{3}{8}}}
{\gam{\frac{1}{8}} \gam{\frac{3}{8}} {(n!)^2}}} \nonumber \\
& \times & \biggl\{{\Psi\left({n+\frac{1}{8}}\right)
 + \Psi\left({n+\frac{3}{8}}\right) -2\Psi\left({n+1}\right)}\biggr\}
 {z^n} \nonumber \,\,\,.
\eeqs
Then the mirror map $t$ valid in the range $|{\psi}|>1$ and
$0 \leq arg \,\,{\psi} < \frac{\pi}{2}$ can be expressed,
\beq
t &=& \frac{1}{2\pi i} \left({\frac{y_2}{y_1} +\sqrt{2} \pi}\right) \nonumber
\\
 &=& \frac{1}{2\pi i} \biggl[{\sqrt{2} \pi -4\log \psi +
\frac{{\thyp{2}{1}} \!\left({\frac{1}{8},\frac{3}{8},1;{\psi}^{-4}}\right)}
{\hyp{2}{1} \!\left({\frac{1}{8},\frac{3}{8},1;{\psi}^{-4}}\right)}
}\biggr] \,\,\,. \label{eqn:tleq1}
\eeq
Also in the range
$|{\psi}| \leq 1$ and $0 \leq arg \,\,{\psi} < \frac{\pi}{2}$,
this map $t$ can be written by the analytic continuation,
\beq
t &=& -\frac{1}{2} +\frac{i}{2} \cdot {\displaystyle
\frac{{Z_1}-{e^{\frac{3}{4} \pi i}} {\left({\tan \frac{\pi}{8}}\right)}^{-1}
\cdot {Z_2}}
{{Z_1}+{e^{\frac{3}{4} \pi i}} {\left({\tan \frac{\pi}{8}}\right)}^{-1}
\cdot {Z_2}}} \,\,\,, \label{eqn:tgeq1} \\
{Z_1} &:= & {\frac{{\left[{\gamm{1}{8}}\right]}^2}{\gam{\frac{3}{4}}}}
 \cdot \hyp{2}{1} \biggl({\frac{1}{8},\frac{1}{8},
\frac{3}{4};{\psi}^{4}}\biggr) \,\,\,, \nonumber \\
{Z_2} &:= & \frac{{\left[{\gam{\frac{3}{8}}}\right]}^2}
{\gam{\frac{5}{4}}} \cdot \psi \,\,
\hyp{2}{1} \biggl({\frac{3}{8},\frac{3}{8},
\frac{5}{4};{\psi}^{4}}\biggr) \,\,\,. \nonumber
\eeq
Moreover the expression of $t$ reads for $|{\psi -1}| \sim 0$,
\beq
t &=& \frac{i}{\sqrt{2}} \cdot {\displaystyle
\frac{{U_1}+{\left({\tan \frac{\pi}{8}}\right)}^{-1}
\cdot {U_2}}
{{U_1}-{\left({\tan \frac{\pi}{8}}\right)}^{-1}
\cdot {U_2}}} \,\,\,, \label{eqn:teq1} \\
{U_1} &:=&  {\frac{{\left[{\gamm{1}{8}}\right]}^{2}}{\gam{\frac{1}{2}}}}
\cdot \hyp{2}{1} \biggl({\frac{1}{8},\frac{1}{8},
\frac{1}{2};1-{\psi}^{4}}\biggr) \,\,\,, \nonumber \\
{U_2} &:=& \frac{{\left[{\gam{\frac{5}{8}}}\right]}^{2}}{\gam{\frac{3}{2}}}
\cdot {({{\psi}^{4} -1})}^{1/2}
\hyp{2}{1} \biggl({\frac{5}{8},\frac{5}{8},
\frac{3}{2};1-{\psi}^{4}}\biggr) \,\,\,. \nonumber
\eeq
The map $t$ connects the complex structure of the mirror manifold ${\mw}$
to the {\kae} structure of the original {\mfd} ${\mm}$. The properties
of $t$ concerning monodromies are postponed till section $5$.

\section{Yukawa coupling}
\pr
In this section, we calculate a two-point correlation
function of (co)homology
classes associated with the {\kae} structure of the original K3 surface
or with the complex structure of its mirror surface.
The notion of the Yukawa coupling is easily extended to d-dimensional case.
This can be regarded as the d-point correlation function of
(co)homology classes correspond to the complex or the {\kae} structure.

\subsection[Yukawa coupling]
{{\kae} potential and metric on the moduli space}
\pr

A metric of the complex moduli space in the mirror ${\mw}$ can be
calculated from a {\kae} potential $K$ {\cite{CO}}.
The {\kae} potential in two-dimension is the same form
as in three-dimensional case;
\beqs
{e^{-K}} &:=&  {\int}_{\mw} {\Om} \wedge \bar{\Om} \\
&=& {{\vpi}_0} {{\bar{\vpi}}_2}
 +{{\vpi}_2} {{\bar{\vpi}}_{0}}
 -2{{\vpi}_1} {{\bar{\vpi}}_{1}} \\
&=& {\biggl|{\frac{1}{{z \cdot {({z-1})}^{1/2}}}
\left({\frac{dz}{dt}}\right)}\biggr|}^{2}
 {(t-\bar{t})}^{2} \\
&=& {{\vpi}_0}{{\bar{\vpi}}_0} \,\, {(t-\bar{t})}^{2} \,\,\,,
\eeqs
where the symbol "{\={\mbox{}}}" represents a complex conjugate.
Also the metric of the complex moduli can be expressed,
\beq
{G_{\psi \bar{\psi}}} &=& \frac{{{\del}^{2}}K}
{\del \psi \del \bar{\psi}} \nonumber \\
&=& \frac{1}{2 {({\Im} t)}^2}
{\left|{\frac{dt}{d \psi}}\right|}^{2} \label{eqn:zam} \,\,\,.
\eeq

\subsection[Yukawa coupling]{Yukawa coupling}
\pr
Two-point coupling ${\kappa}_{\psi \psi}$ of the (co)homology class
 ${{\cal H}^{2}} ({\mw})$ is defined as,
\begin{equation}
{\kappa}_{\psi \psi} := -{\int}_{\mw}\Omega \wedge
{\chi}\sb{\psi}\sp{\mu}\wedge{\chi}\sb{\psi}\sp{\nu}\,
{\Omega}_{\mu \nu}\;\;,
\end{equation}
where
\beq
{\chi}\sb{\psi}\sp{\mu} &:=& \frac{1}{2{\Vert\Omega\Vert}^2}
\bar{\Omega}^{\mu \nu}{\chi}_{\psi \rho \bar{\nu}}\, dx^{\bar{\nu}}\,\,\,,\\
{\chi}_{\psi} &:=& {\chi}_{\psi \rho \bar{\nu}}dx^{\rho}\wedge
dx^{\bar{\nu}}
\in{\cal H}^{1,1}(\mw)\,\,\,,
\eeq
with the help of the Kodaira's theorem; ${\dis \frac{d\Omega}{d\psi}}
={\kappa}_{\psi}{\Omega}+{\chi}_{\psi}\in{\cal H}^{2,0}(\mw)
{\bigoplus}{\cal H}^{1,1}(\mw)$.
The above formula is reduced to
\beq
{\kappa}_{\psi \psi} &:=& -{{\int}_{\mw}} \Om \wedge
\frac{{d^2} \Om}{d {{\psi}^2}} \\ \nonumber
&=& 2 \biggl\{{{{\left({\frac{d {{\vpi}_1}}{d \psi}}\right)}^{2}}
 -{\left({\frac{d {{\vpi}_0}}{d \psi}}\right)} \cdot
 {\left({\frac{d {{\vpi}_2}}{d \psi}}\right)}}\biggr\} \\ \nonumber
&=& 2 {\left[{\frac{-4\psi}{{(1-{\psi}^{4})}^{1/2}}}\right]}^{2} \\ \nonumber
&=& \frac{32{{\psi}^{2}}}{1-{{\psi}^{4}}} \,\,\,.
\eeq
Here we have to remember
that the holomorphic $(2,0)$-form takes value in the
holomorphic line bundle.
We have arbitrariness
in the choice of $\Om$, i.e. a guage symmetry in physical speaking,
\[
{\Om}(\psi) \rightarrow f(\psi) {\Om}(\psi) \,\,\,,
\]
where $f(\psi)$ is an arbitrary holomorphic function.
Under this transformation, the {\kae} potential $K$ turns into
$K-\log f(\psi)- \log \bar{f} ({\bar{\psi}})$.
By the aid of the Kodaira's theorem,
the two point coupling ${\kappa}_{\psi \psi}$ becomes
${{\left({f(\psi)}\right)}^2} {\kappa}_{\psi \psi}$.

To relate with the Yukawa couping of the {\kae}
structure, we convert the coordinate $\psi$
to the one $t$ given by the
mirror map.
We fix gauge arbitrariness by choosing
the {\kae} potential to be a simple form,
\[
K=- \log {(t-\bar{t})}^{2} \,\,\,.
\]
In turns in order to get this formula,
we have to perform a gauge transformation
on the original $(2,0)$-form $\Om$,
\[
\Om \rightarrow \frac{1}{{{\vpi}_0}({\psi})} \,\, \Om \,\,\,.
\]
After these transformation, the coupling ${\kappa}_{\psi \psi}$ becomes
${\kappa}_{\psi \psi}\left(\frac{d\psi}{dt}\right) \times
\frac{1}{{({{\vpi}_{0}})}^2}$.
The two point coupling ${\kappa}_{tt}$ associated with the {\kae}
structure of the {\mfd} ${\mm}$ becomes
\beq
{{\kappa}_{tt}} &=& {{\kappa}_{\psi \psi}}
{{\left({\frac{d \psi}{dt}}\right)}^{2}} \!
\times \frac{1}{{({{\vpi}_{0}})}^2} \nonumber \\
&=& 2 \,{{\biggl[{\,\, \frac{-4 \psi}{{(1-{{\psi}^{4}})}^{1/2}}
 {\left({\frac{dt}{d \psi}}\right)}^{-1} \,\,}\biggr]}^{2}}
\times \frac{1}{{({{\vpi}_{0}})}^2} \nonumber \\
&=& 2 \,\, {{({{\vpi}_{0}})}^{2}}
 \times \frac{1}{{({{\vpi}_{0}})}^2}  \nonumber \\
&=& 2 \label{eqn:kappa} \,\,\,.
\eeq
{}From {\cite{DSWW,DG}}, it is shown that
\beq
{\kappa}_{tt} &=& \int_{\mm} e{\wedge}e+
(instanton\; corrections)\;\;,\\
e & \in & {\cal H}^{1,1}(\mm)\;\;.
\eeq
Moreover the value of $\int_{\mm}e{\wedge}e$ is $2$.

In contrast with the quintic case {\cite{MIR4}},
the coupling in the {\kae} {\md}
contains no quantum (instanton) correction. Coefficients of correction terms,
if exist, are interpreted as the number of rational curves
embedded in this {\mfd} {\cite{MIR4,RC,WIT}}.
Therefore our calculation shows that the
K3 surface ${{p}_{\psi}}=0$ contains no rational curve on it.

Let us consider this more Algebraic geometricaly.
For simplicity we consider rational curves
with degree one. A Riemann surface with genus $0$ $i.e.$ ${\bf CP}^1$
is parametrized by
a projective coordinate $(u,v)$.
If a rational curve on the K3 surface exists,
there exists a set of complex numbers $({{a}_{i},{b}_{i}})
 \hspace{2mm} (i=1,2,3,4)$ such that the point
$({{{a}_{1}}u+{{b}_{1}}v,{{a}_{2}}u+{{b}_{2}}v,
{{a}_{3}}u+{{b}_{3}}v,{{a}_{4}}u+{{b}_{4}}v})$ in ${\bf CP}^3$
is on the defining equation for arbitrary sets of $(u,v)$.
In short, the relation
should be satisfied,
\beqs
&& {({{{a}_{1}}u+{{b}_{1}}v})}^{4}
+{({{{a}_{2}}u+{{b}_{2}}v})}^{4}
+{({{{a}_{3}}u+{{b}_{3}}v})}^{4}
+{({{{a}_{4}}u+{{b}_{4}}v})}^{4} \\
&& =4 \psi ({{{a}_{1}}u+{{b}_{1}}v})
({{{a}_{2}}u+{{b}_{2}}v})
({{{a}_{3}}u+{{b}_{3}}v})
({{{a}_{4}}u+{{b}_{4}}v}) \,\,\,,
\eeqs
for all values of $(u,v)$.
In this case, there is five relations.
But there exists one automorphism,
\[
(u,v) \rightarrow ({{\hat{A} u+\hat{B} v},{\hat{C} u+\hat{D} v}}) \,\,\,.
\]
Collecting these, one can calculate the number of the freedom that
the rational curve has,
\[
8-5-4=-1\,\,.
\]
It is overdetermined.
Therefore there exists no rational curve with degree one and
we exhibit no quantum correction exists in the coupling
of the K3 surface.

A prepotential ${\cal F}^{(t)}$
of ${\kappa}_{tt}$ (\ref{eqn:kappa})
can be defined in this gauge,
\[
{{\cal F}^{(t)}}={t^2} \,\,\,.
\]
On the other hand in the old gauge, this prepotential is written
by taking account of the fact that the prepotentaial is a
homogeneous polynomial with degree {$2$},
\[
{{\cal F}^{({\vpi})}}={{({{\vpi}_0})}^{2}} {t^2} ={{\vpi}_0}{{\vpi}_2}
={{({{\vpi}_1})}^2} \,\,\,.
\]
Also the differential operator ${\del}_{t}$ acting on ${\cal F}^{(t)}$ is
replaced by a covariant derivative for degree $2$,
which we introduce as follows;
\beqs
{\cal D}\sb{t}\sp{(2)} &=& {{({{\vpi}_0})}^2} {{\del}_{t}}\,
{{({{\vpi}_0})}^{-2}} \\
&=& {{\del}_{t}}-2\,\frac{{\del}_{t}{{\vpi}_0}}{{\vpi}_0} \,\,\,,
\eeqs
such that
\beqs
{{\cal D}_{t}}{{\cal D}_{t}} {{\cal F}^{({\vpi})}} =2 {{({{\vpi}_0})}^2}
\,\,\,,
\eeqs
is the coupling obtained in the old gauge.

\section{Monodromy}
\pr
In this section, we study monodromy transformations and give monodromy
matrices, fundamental regions.

\subsection[Monodromy]{Monodromy}
\pr
The defining equation of one parameter family of K3 surface was represented
in (\ref{eqn:de}). If the conditions
${\del}p/{\del}{X_i}=0,\hspace{2mm} (i=1,2,3,4)$ are
satisfied for some $\psi$, this variety is singular.
These conditions are rewritten,
\[
{{({{X_1}{X_2}{X_3}{X_4}})}^{3}} {({{{\psi}^4}}-1)}=0 \,\,\,,
\]
and this variety is singular at ${{\psi}^4}=1, {\infty}$,
$i.e.\,{\psi} ={{\alpha}^n},{\infty}, \hspace{2mm}
({{\alpha}={e^{\frac{\pi}{2} i}}=i\,;\,\,n=0,1,2,3})$.
If one rotates the values of the parameter $\psi$ around the special value
analytically in the complex $\psi$-plane, homology cycles ${\gamma}_i$
of the variety (\ref{eqn:de}) turn into linear combinations of other
homology cycles ${\dis{\sum}_{j}} c_{ij} {\gamma}_j$.
As a consequence, the periods
also change into linear combinations of others. The matrices associated
with this linear transformation are called monodromy matrices and contain
geometrical information on the variety.

\subsection[Monodromy]{Monodromy matrices}
\pr
In order to obtain the monodromy matrices, let us consider two monodromy
transformations ${\ca},{\ct}$ {\cite{MIR4,MOD}},
which act on the mirror map $t$. The ${\ca}$
is a transformation of $t({\psi})$ associated with the rotation around
the point ${\psi}=0$ with an angle $\frac{\pi}{2}$ counterclockwise in the
complex $\psi$-plane, $i.e.\, \psi \rightarrow {e^{\frac{\pi}{2} i}} \psi $.
Also the ${\ct}$ is associated with the rotation around
the point ${\psi}=1$ with angle $2\pi$ counterclockwise,
$i.e.\, \psi =1+\epsilon \rightarrow \psi =1+\epsilon {e^{2\pi i}}
\hspace{2mm} (\epsilon \,;\, \mbox{small complex number}) $.
As is well-known, one can express the monodromy transformation
${{\ct}_m} \hspace{2mm} (m=0,1,2,3,\infty ) $ {\cite{MIR4,LER}},
which are associated with
the rotations around the points
$\psi ={{\alpha}^0},{{\alpha}^1},{{\alpha}^2},{{\alpha}^3},\infty
\hspace{2mm} ({{\alpha}={e^{\frac{\pi}{2} i}}=i})$ with the angle $2\pi$
respectively,
\beq
{{\ct}_m} &=& {{\ca}^m}{\ct}{{\ca}^{-m}} \,\,\,,\hspace{3mm}
(m=0,1,2,3) \,\,\,,\label{eqn:tm} \,\,\,\ \\
{{\ct}_{\infty}} &=&
{\left({{{\ct}_0}{{\ct}_1}{{\ct}_2}{{\ct}_3}}\right)}^{-1}
= {\left({{\ct}{\ca}}\right)}^{-4} \label{eqn:tf} \,\,\,.
\eeq
Also the monodromy transformation ${\hat{\ct}}_\infty $ associated with the
rotation around $\psi =\infty$ with an angle $\frac{\pi}{2}$ is
represented as,
\[
{{\hat{\ct}}_{\infty}}= {\left({{\ct}{\ca}}\right)}^{-1} \,\,\,.
\]
Firstly let us consider the actions of ${\ca}$ on ${Z_1},\,{Z_2}$
in equation (\ref{eqn:tleq1}),
\beqs
{\ca} {Z_1}({\psi}) &=& {Z_1}({\psi}) \,\,\,,\\
{\ca} {Z_2}({\psi}) &=& {\alpha}{Z_2}({\psi}) \,\,\,.
\eeqs
We obtain the relation,
\beq
{\ca}t(\psi) &=& \frac{\sqrt{2} t({\psi})+\frac{1}{\sqrt{2}}}
{-\sqrt{2} t({\psi})} \nonumber \\
&=& -1-\frac{1}{2t({\psi})} \label{eqn:at} \,\,\,.
\eeq
This transformation (\ref{eqn:at}) is a composition of an inversion
$-\frac{1}{2t({\psi})}$ and a translation $t({\psi})-1$.
We get monodromy matrices ${\tilde{\ia}},\,{\ia}$ associated with ${\ca}$
on the bases ${\mbox{}^t}({\vvp{0}},{\vvp{1}},{\vvp{2}})$,
${\mbox{}^t}({\om{0}},{\om{1}},{\om{2}})$, respectively,
\begin{equation}
\tilde{\ia} =\left(
\begin{array}{rrr}
0 & 0 & 2 \\
0 & -1 & -2 \\
\frac{1}{2} & 2 & 2
\end{array}
\right) \nonumber
\,\,\,,\\
{\ia} =\left(
\begin{array}{rrr}
\frac{9}{4} & 1 & -\frac{7}{4} \\
-2 & -1 & 2 \\
-\frac{1}{4} & -1 & -\frac{1}{4}
\end{array}
\right) \nonumber
\,\,\,.
\end{equation}
Secondly the transformation ${\ct}$ should be considered.
Because of the actions of ${\ct}$ on ${U_1},\,{U_2}$ in the
equation (\ref{eqn:teq1}),
\beqs
{\ct}{U_1}({\psi}) &=& {U_1}({\psi}) \,\,\,,\\
{\ct}{U_2}({\psi}) &=& -{U_2}({\psi}) \,\,\,,
\eeqs
the relation is obtained,
\beq
{\ct}t({\psi})=-\frac{1}{2t({\psi})} \label{eqn:tt} \,\,\,.
\eeq
This is an inversion of $t({\psi})$.
Monodromy matrices $\tilde{\itt},\,{\itt}$ on the bases
${\mbox{}^t}({\vvp{0}},{\vvp{1}},{\vvp{2}})$,
${\mbox{}^t}({\om{0}},{\om{1}},{\om{2}})$ associated with ${\ct}$
are given,
\begin{equation}
\tilde{\itt} =\left(
\begin{array}{rrr}
0 & 0 & 2 \\
0 & -1 & 0 \\
\frac{1}{2} & 0 & 0
\end{array}
\right) \nonumber \,\,\,,\\
{\itt} =\left(
\begin{array}{rrr}
\frac{5}{4} & 0 & -\frac{3}{4} \\
0 & -1 & 0 \\
\frac{3}{4} & 0 & -\frac{5}{4}
\end{array}
\right) \nonumber \,\,\,.
\end{equation}
Collecting the above considerations, we list the results.
Monodromy matrices on the basis ${\mbox{}^t}({\vvp{0}},{\vvp{1}},{\vvp{2}})$
are expressed,
\begin{equation}
\tilde{\ia} =\left(
\begin{array}{rrr}
0 & 0 & 2 \\
0 & -1 & -2 \\
\frac{1}{2} & 2 & 2
\end{array}
\right) \nonumber \,\,\,,\,\,\,\,\,\,
\tilde{\itt} =\left(
\begin{array}{rrr}
0 & 0 & 2 \\
0 & -1 & 0 \\
\frac{1}{2} & 0 & 0
\end{array}
\right) \nonumber \,\,\,,\\
\end{equation}
\begin{equation}
{{\tilde{\itt}}_{1}} =\left(
\begin{array}{rrr}
2 & 12 & 18 \\
-1 & -5 & -6 \\
\frac{1}{2} & 2 & 2
\end{array}
\right) \nonumber \,\,\,,\,\,\,\,\,\,
{{\tilde{\itt}}_2} =\left(
\begin{array}{rrr}
0 & 0 & 2 \\
0 & -1 & -4 \\
\frac{1}{2} & 4 & 8
\end{array}
\right) \nonumber \,\,\,,\\
\end{equation}
\begin{equation}
{{\tilde{\itt}}_{3}} =\left(
\begin{array}{rrr}
2 & 4 & 2 \\
-3 & -5 & -2 \\
\frac{9}{2} & 6 & 2
\end{array}
\right) \nonumber \,\,\,,\,\,\,\,\,\,
{{\tilde{\itt}}_{\infty}} =\left(
\begin{array}{rrr}
1 & 0 & 0 \\
4 & 1 & 0 \\
16 & 8 & 1
\end{array}
\right) \nonumber \,\,\,,\\
\end{equation}
\begin{equation}
{{\widehat{\tilde{\itt}}}_{\infty}} =\left(
\begin{array}{rrr}
1 & 0 & 0 \\
1 & 1 & 0 \\
1 & 2 & 1
\end{array}
\right) \nonumber \,\,\,.
\end{equation}
Also monodromy matrices on the basis
${\mbox{}^t}({\om{0}},{\om{1}},{\om{2}})$ read,
\begin{equation}
{\ia} =\left(
\begin{array}{rrr}
\frac{9}{4} & 1 & -\frac{7}{4} \\
-2 & -1 & 2 \\
-\frac{1}{4} & -1 & -\frac{1}{4}
\end{array}
\right) \nonumber \,\,\,,\,\,\,\,\,\,
{\itt} =\left(
\begin{array}{rrr}
\frac{5}{4} & 0 & -\frac{3}{4} \\
0 & -1 & 0 \\
\frac{3}{4} & 0 & -\frac{5}{4}
\end{array}
\right) \nonumber \,\,\,,\\
\end{equation}
\begin{equation}
{{\itt}_{1}} =\left(
\begin{array}{rrr}
\frac{45}{4} & 7 & -\frac{35}{4} \\
-7 & -5 & 5 \\
\frac{35}{4} & 5 & -\frac{29}{4}
\end{array}
\right) \nonumber \,\,\,,\,\,\,\,\,\,
{{\itt}_2} =\left(
\begin{array}{rrr}
\frac{77}{4} & 18 & -\frac{27}{4} \\
-18 & -17 & 6 \\
\frac{27}{4} & 6 & -\frac{13}{4}
\end{array}
\right) \nonumber \,\,\,,\\
\end{equation}
\begin{equation}
{{\itt}_{3}} =\left(
\begin{array}{rrr}
\frac{21}{4} & 5 & \frac{5}{4} \\
-5 & -5 & -1 \\
-\frac{5}{4} & -1 & -\frac{5}{4}
\end{array}
\right) \nonumber \,\,\,,\,\,\,\,\,\,
{{\itt}_{\infty}} =\left(
\begin{array}{rrr}
9 & 4 & 8 \\
4 & 1 & 4 \\
-8 & -4 & -7
\end{array}
\right) \nonumber \,\,\,,\\
\end{equation}
\begin{equation}
{{\widehat{\itt}}_{\infty}} =\left(
\begin{array}{rrr}
\frac{3}{2} & 1 & \frac{1}{2} \\
1 & 1 & 1 \\
-\frac{1}{2} & -1 & \frac{1}{2}
\end{array}
\right) \nonumber \,\,\,.
\end{equation}
All these matrices belong to the group $SO(2,1)$.

\subsection[Monodromy]{Fundamental region}
\pr
Recall the result (\ref{eqn:zam}) in section $4$,
we obtain the metric of the {\kae} moduli space,
\[
{G_{t \bar{t}}}=\frac{1}{2{{({\Im}t)}^2}} \,\,\,.
\]
A Ricci tensor and a scalar curvature are calculated,
\beqs
{R_{t \bar{t}}}&=& \frac{-1/2}{{({{\Im}t})}^2} \,\,\,,\\
R &=& -4 \,\,\,.
\eeqs
It is sufficient to consider the upper half $t$-plane in considering the
{\md} because the curvature is a negative constant number.
The monodromy transformations are generated by two generators
${\ca},\,{\ct}$ or by other ones ${\ca},\,{{\widehat{\ct}}_{\infty}}$.
These act on $t$,
\beqs
{{\widehat{\ct}}_{\infty}} t({\psi}) &=& t({\psi})+1 \,\,\,,\\
{\ca}\,\,t({\psi}) &=& -1-\frac{1}{2t({\psi})} \,\,\,.
\eeqs
The upper half $t$-plane is mapped to itself and is covered once and only
once by combinations of these actions. A fundamental region is drawn
by considering these expressions in Fig.1.
The fixed point of ${\ca}$ is $t=-\frac{1}{2} +\frac{i}{2}$ and that of
${\ct}$ is $t=\frac{i}{\sqrt{2}}$.
These correspond to the values of $t$ at ${\psi}=0$, ${\psi}=1$ respectively.
On the contrary, the infinity $t= \infty$ is the fixed point of
${\widehat{\ct}}_{\infty}$ and corresponds to the point ${\psi}=\infty $.
The $t$ with large imaginary part in the fundamental
region is mapped to some
point with small imaginary part in some region by the action of ${\ca}$.
In short, large ${\Im}t$ region is associated with small ${\Im}t$ region
by modular transformation.
That is suggestive of some kind of duality.

Also one can take
another function in order to study monodromy group $SL(2,{\bf R})$ by
the standard method {\cite{KF}}.
Firstly, let us introduce a function $\gamma ({\psi})$,
\beq
{\gamma}({\psi}) &:=& i \; {\displaystyle
\frac{\zg{1}-{e^{\frac{3}{4} \pi i}} \zg{2}}
{\zg{1}+{e^{\frac{3}{4} \pi i}} \zg{2}}} \,\,\,,\\
&=& \frac{i}{\pi} \left({\tan \frac{3\pi}{8}}\right)
\Biggl\{ \log {\psi}^4 -\pi i \nonumber \\
&+& \biggl[ \,\, {\fsum{n=0}}{\displaystyle \frac{\gam{n+\frac{3}{8}}
\gam{n+\frac{5}{8}}}
 {{{(n!)}^{2}}{{\psi}^{4n}}}}
\biggl\{{2\Psi (n+1)-\Psi \left({n+\frac{3}{8}}\right) -\Psi
\left({n+\frac{5}{8}}\right) }\biggr\} \biggr] \nonumber \\
&\times & {\biggl[ \,\, {\fsum{n=0}}{\displaystyle \frac{\gam{n+\frac{3}{8}}
\gam{n+\frac{5}{8}}}
{{{(n!)}^{2}}{{\psi}^{4n}}}} \,\, \biggr]}^{-1} \, \Biggr\} \,\,\,,\\
\zg{1} &:=& {\displaystyle \frac{{\left[{\,\gamm{3}{8}}\right]}^2}
{\gam{\frac{3}{4}}}} \cdot \hyp{2}{1}
\biggl({\frac{3}{8},\frac{3}{8},\frac{3}{4}
;{{\psi}^{4}}}\biggr) \nonumber \,\,\,,\\
\zg{2} &:=& {\displaystyle \frac{{\left[{\,\gamm{5}{8}}\right]}^2}
{\gam{\frac{5}{4}}}} \cdot \, \psi \,\,\hyp{2}{1}
\biggl({\frac{5}{8},\frac{5}{8},\frac{5}{4}
;{{\psi}^{4}}}\biggr) \nonumber \,\,\,.
\eeq
The $\gamma$ is a map of the ${\psi}^{4}$-plane to a pair of triangles in the
upper half $\gamma$-plane, which together constitute a fundamental region of
this transformation in Fig.2. On this $\gamma$, the transformations
${\ca},\,{\ct},\,{\widehat{\ct}_{\infty}}$ are represented respectively,
\begin{equation}
{{\ia}^{\gamma}}=\left(
\begin{array}{rr}
-\frac{1}{\sqrt{2}} & -\frac{1}{\sqrt{2}} \\
\frac{1}{\sqrt{2}} & -\frac{1}{\sqrt{2}}
\end{array}
\right) \,\,\,,\,\,\,\,
{{\itt}^{\gamma}}=\left(
\begin{array}{rr}
-\frac{1}{\sqrt{2}} & 2+ \frac{3}{\sqrt{2}} \\
-\frac{1}{\sqrt{2}} & 2+ \frac{1}{\sqrt{2}}
\end{array}
\right) \,\,\,,
\end{equation}
\begin{equation}
{\widehat{\itt}}^{\gamma}_{\infty} =\left(
\begin{array}{rc}
1 & 2({\sqrt{2} +1}) \\
0 & 1
\end{array}
\right) \,\,\,.
\end{equation}
Fixed points of these transformations are ${\gamma}(0)=i$,
${\gamma}(1)= \sqrt{2} +1$, and ${\gamma}({\infty})=i{\infty}$
respectively.

\section{Torus}
\pr
We take a Torus which is represented by the following polynomial
in ${\bf CP}^2$;
$$
{\cal M};\, p={X_1^3}+{X_2^3}+{X_3^3}-3{\psi}{X_1}{X_2}{X_3}=0\,\,\,.
$$
The Hodge diamond of torus is known,
$$
\begin{array}{ccccc}
 & 1 &  \\
1 & & 1 \\
 & 1 &\,\,\,\,\,.
\end{array}
$$
To obtain the mirror manifold,
we divide ${\cal M}$ by a discrete symmetry $G=Z_3$ ;
\begin{eqnarray}
& (1,0,2) & \,\,\,,\\
& (0,1,2) & \,\,\,,
\end{eqnarray}
where the symbol $(n_1,n_2,n_3)$ means the transformation,
\begin{eqnarray}
(X_1,X_2,X_3) & {\rightarrow} & ({\alpha}^{n_1}X_1,{\alpha}^{n_2}X_2,
{\alpha}^{n_3}X_3)\,\,\,, \\
 {\alpha} & := & e^{2{\pi}i/3}\,\,\,.
\end{eqnarray}

Let {$\mw$} be the mirror manifold. Euler number of ${\mw}$ is ;
$$
{\chi}({\mw})={\frac{0-0}3}+0=0\,\,\,.
$$
Again torus is self-dual. We can calculate the periods on torus
exactly
with no help of Picard-Fuchs equation.
Let ${\alpha},{\beta}$  be the duals of $A$-cycle and $B$-cycle
respectively.
Then a holomorphic $(1,0)$-form ${\Omega}$ on the torus is decomposed
with the basis ${\alpha}, {\beta}$ into
$$
{\Omega}={\omega}_0{\alpha}-{\omega}_1{\beta}\,\,,
$$
where
\begin{eqnarray}
&&{\int_{\mw}}{\alpha}{\wedge}{\beta}=1\,\,\,\\
&&{\int_{\mw}}{\alpha}{\wedge}{\alpha}=
  {\int_{\mw}}{\beta}{\wedge}{\beta}=0 \,\,\,,\\
{\alpha},{\beta}\,{\in}\,{\cal H}_1({\mw})\;,\;\;\;\;
{\Omega}\,{\in}\,{\cal H}^{1,0}({\mw})\,\,\,,\\
{\omega}_0={\int_A}{\Omega}
 =-\frac{1}{3{\psi}}(2{\pi}i)^3{}_2F_1(\frac{1}{3},\frac{2}{3};1;
{\psi}^{-3})\,\,, \\
{\omega}_1={\int_B}{\Omega}
=-\frac{1}{3\psi}{(2{\pi}i)^3}\Biggl[{}_2F_1(\frac{1}{3},\frac{2}{3};1;
{\psi}^{-3})\log {{\psi}^{-3}}+{}_2{\tilde{F}}_1(\frac{1}{3},\frac{2}{3};1;
{\psi}^{-3})\Biggr]\,\,\,.
\end{eqnarray}
Also mirror map is defined,
$$
t:=\frac{1}{2\pi i}\frac{\omega_1}{\omega_0}\;\;\; .
$$
It is a standard coordinate of the complex moduli of the torus.
Corresponding to generators of the modular transformations,
$t$ transforms as  follows ;
\beq
&&\left\{
\matrix{
A & \longrightarrow & A+B \cr
B & \longrightarrow & B \cr
}
\right.
\;\;\;
\left\{
\matrix{
A & \longrightarrow & A \cr
B & \longrightarrow & B+A \cr
}
\right. \nonumber \\
&& t \longrightarrow \frac{t}{1+t}
\;\;\;\;\;\;\;\;\;\;\;\;\;\;\;\;\;\;\;\;\;
t \longrightarrow t+1 \nonumber
\eeq

These generate $SL(2;{\bf Z})$, which is widely known.
The interesting object is one-point coupling ;
\beq
{\kappa}_t & = & \int_{\mw}e \\
& = & -\frac{1}{{\omega_0}^2}\times \int_{\mw}
\Omega\wedge\frac{d\Omega}{dt}  \nonumber \\
          & = & 1  \;\; .\nonumber
\eeq
The factor in the right-hand side at the first line of the above equations
$1/{{\omega_0}^2}$ is chosed when we fix the gauge, and it is known as
the primitive factor in Algebraic geometry.
We conclude there is no instanton correction
in accordance with Algebraic geometry.
Similarly,prepotential and covariant derivative can be constructed.

\section{Conclusion}
\pr
In this paper, we dealt with the K3 surface mainly. The coupling of
(co)homology elements is calculated under the some gauge choice.
In this gauge, the {\kae} potential becomes simple form in the variables
$t,\,\bar{t}$.
The two point coupling becomes constant number and
no quantum correction exists.
In the mathematical language, this fact about the coupling means that
there is no rational curve on the K3 surface in contrast with the quintic
hypersurface.
The monodromy group acts on the mirror map $t$ as the $SL(2,{\bf R})$ or
on the periods as the $SO(2,1)$.
There is no modification in this monodromy group owing to the quantum
corrections. This property seems to stem from the fact that the mirror map in
this case can be identified with a solution of the schwarzian equation
(\ref{eqn:sch}) by using a $PSL(2)$ symmetry (\ref{eqn:arb}).
In the quintic case, the periods cannot be expressed
in the similar simple form and the monodromy group acting on the mirror map
is not the $SL(2,{\bf R})$ because of the quantum corrections.

\section*{Acknowledgement}
\pr
The authors greatly thank Prof.~T.~Eguchi for helpful discussions
and kindful encouragement. They also thank M.~Jinzenji for useful
discussions.

\end{document}